%% file: main.tex
\documentclass[conference]{IEEEtran}
\usepackage{cite}

\ifCLASSINFOpdf
   \usepackage[pdftex]{graphicx}
\else
\fi
\usepackage[cmex10]{amsmath}
\usepackage{array}

\usepackage[tight,footnotesize]{subfigure}
\usepackage[hyphens]{url}

\usepackage{multirow}

\begin{document}
%
\title{\#mytweet via Instagram: Exploring User Behaviour across Multiple Social Networks}


\author{
	Bang Hui Lim$^{1}$\qquad 
	Dongyuan Lu$^{1}$\qquad
	Tao Chen$^{1}$\qquad 
	Min-Yen Kan$^{1,2}$\\ 
	$^{1}$School of Computing, National University of Singapore\\
	$^{2}$NUS Interactive and Digital Media Institute, Singapore\\
	limbanghui@u.nus.edu.sg \quad \{ludy,taochen,kanmy\}@comp.nus.edu.sg 
}

\maketitle
\begin{abstract}
We study how users of multiple online social networks (OSNs) employ
and share information by studying a common user pool that use six OSNs
-- Flickr, Google+, Instagram, Tumblr, Twitter, and YouTube. We
analyze the temporal and topical signature of users' sharing
behaviour, showing how they exhibit distinct behaviorial patterns on
different networks.  We also examine \textit{cross-sharing} ({\it
  i.e.}, the act of user broadcasting their activity to multiple OSNs
near-simultaneously), a previously-unstudied behaviour and demonstrate
how certain OSNs play the roles of originating source and destination
sinks.

\end{abstract}


%
\IEEEpeerreviewmaketitle

\input{1_introduction}
\input{2_relatedwork}

\input{3_dataset}
\input{4_multi_network}
\input{5_cross_network}
\input{6_conclusion}




%


\bibliographystyle{IEEEtran}
\bibliography{IEEEabrv,bibsource}

\end{document}

%% file: 1_introduction.tex
\section{Introduction}
\label{section:1}

Reading and posting from social networks has become a staple part of
daily activities for many. The Pew Internet Project's January 2014
research reported that 74\% of online adults use social networking
sites\footnote{\url{http://www.pewinternet.org/fact-sheets/social-networking-fact-sheet/}}. More recently, the Global Web Index's January 2015 Social
Report monitored engagement with close to 50 online social networks
(OSN) observed an average of 5.54 social media accounts per user, with
2.82 being used 
actively\footnote{\url{http://www.globalwebindex.net/blog/internet-users-have-average-of-5-social-media-accounts}}.  

Even with so many social network activity and the wide variety of
networks, the published literature reports little about their usage.
We know precious little about what users actually do on OSNs, aside
from our own individual use.  Even less is known about how individuals
interact across multiple OSNs.  Many functionalities across networks
are similar.  So why do people find themselves using more than one?
Are what individuals do on one network the same as their behavior on
another?  Does participation in one network impact their activity on
another?  

These questions are important, but yet remain unaddressed by existing
literature.  With many of today's users being engaged on multiple
platforms, do studies limited to individual social network platforms
provide a good picture of user behaviour in general?  To date, there
has not been a definitive answer to this question.

The above are all questions about single networks, but users can use
several OSNs simultaneously.  One artifact of this is {\it
  cross-network sharing} (akin cross-posting in mailing lists) -- when
people post about their activities on more than one network.  To our
knowledge, why and how do people do such cross-sharing has not
previously been investigated.

We aim to address both issues of multi- and cross-OSN behavior in part
in this work.  We present an exploratory user-centric study on a large
sample of users that participate in multiple OSNs. In particular, we
analyse over 15,000 individuals that participate and link their
accounts on six platforms -- \textit{Flickr, Google+, Instagram,
  Tumblr, Twitter, and YouTube} -- through their publicly-available
profile descriptions and public sharing activity on these networks.
Through our study of this dataset on macro-, meso- and micro-scale
analyses, we conclude that single network analyses is limited and does
not yield representative holistic patterns.  Even though we only study
public data, we find support for the claim that each network is
different and has a particular social networking niche to fill.

We first review related work in Section~\ref{section:2} and then
describe our dataset in detail in Section~\ref{section:3}.  We then
analyse user profiles, posting time, post topic and professions with
respect to {\it multiple networks} in
Section~\ref{section:4}. Finally, in Section \ref{section:5} we turn
to {\it cross-network} interactions, discovering a clear topology of
source and sink network relationships among public OSNs.

%% file: 2_relatedwork.tex
\section{Related Work}
\label{section:2}
Structure, content and user behavior are three major aspects that can
be said to characterize research on social networks.  Leveraging the
user relationship graph, early studies
\cite{ahn2007www,mislove2007imc} investigated the structural
properties among networks such as Flickr, YouTube and Myspace.  They
confirmed several empirical observations true of other netowkrs: that
they follow a power law distribution and exhibit both small-world and
scale-free properties.  More recent work examined how the user
generated content can be leveraged for knowledge discovery and by
examining networks over time, topic evolution.
For example, Althoff {\it et al.} presented a
comprehensive study about the evolution of topics across three online
media streams~\cite{althoff2013mm}.

The third aspect, user behavior, plays a key role.  Understanding user
behavior is a key modeling problem as it affects the social network
structure as well as attempts to best model users themselves.  For
example, Lerman {\it et al.} conducted empirical analyses of
user activities on Digg and Twitter to assess how it affected
dissemination patterns~\cite{lerman2010icwsm}.  Other user activities, such as social
connection \cite{leskovec2008kdd}, content generation
\cite{guo2009kdd}, participation in conversation \cite{de2011wsm}, 
have also been studied to gain better understanding of OSNs.

However, a major shortcoming of these works is that they are limited
to macroscopic analyses from a network (graph) perspective.  Most
works have neglected any analysis of micro- (individuals) or meso-
(small aggregrates) levels.  In contrast, we perform a user-centric
analysis: following the same users across their multiple social
networks to uncover how and why users participate in and interact
among their multiple social networks.

To understand how user interact with multiple OSNs, early studies
exploited user clickstream data from passively monitored network
traffic \cite{schneider2009IMCunderstanding} or a social network
aggregator \cite{benevenuto2009IMCcharacterizing}.  However,
clickstream data is difficult to obtain for cross-network analysis due
to privacy reasons, rendering such techniques difficult to execute in
practice.
While user generated content in part overcomes these problems (unlike web search history, user generated content in OSNs is often created by users voluntarily and for public consumption, alleviating privacy concerns), the additional problem of user linkage needs to be solved.  Automated user linkage aims to link user accounts among different social networks that belong to the same person in the real world~\cite{zafarani2013KDDconnecting,Zhang2014}. 

Given the associated user accounts and collected user-generated content (UGC) among different networks, works can then address the subsequent analysis and build downstream application.  For example, Kumar {\it et al.} introduced a pioneer analysis work on users' migration patterns across seven OSNs, which provides guidelines to prevent or encourage social media traffic~\cite{kumar2011understanding}. Chen {\it et al.} examined the extent of personal information revealed by users across multiple OSNs, and found the amount of revealed information in profile is correlated with occupations and the use of pseudonyms~\cite{Chen2012}. 


There are relatively few studies on cross-posting.  Many OSNs provide
a cross-site linking functionality -- linking two accounts on
different platforms so that information can be (automatically) shared
between them. This is often a mandatory step for users to perform
cross-posting.  Chen {\it et al.}, using
a data-driven approach, investigated how users cross-linked their
Foursquare accounts to other OSNs, namely Facebook and Twitter~\cite{chen2014understanding}.  Most
relevant to cross-posting is work by Ottoni {\it et al.}
\cite{ottoni2014pins}.  They studied the correspondence and
discrepancy in user activities across Twitter and Pinterest. Their key
findings are that 1) users often generate content on Pinterest and
then share them to Twitter, and that 2) users exhibit more focused
interests on Pinterest than Twitter.  In contrast, our research is
based on six OSNs that span a more diverse set of media, and
correlates these with analyses from other perspectives (e.g.,
temporal, profession).



%% file: 3_dataset.tex
\section{Dataset}
\label{section:3}


A user-centric study of cross-OSN behavior requires a collective study
of many individual users, each of whom use multiple OSNs.  We
purposefully sidestep the issue of user linkage to focus our attention on user
behavior.

We leverage an OSN aggregation service called {\it
  About.me}\footnote{\url{http://about.me/}}, which enables people to
easily create a public online identity that unifies a self-described
short biography with prominent links to the person's other OSN
accounts and personal websites.  According to web analytics provided
by Alexa
Internet\footnote{\url{http://www.alexa.com/siteinfo/about.me},
  retrieved on 27 April 2015.}, About.me is ranked as the 1,731th most
popular website worldwide, which is more popular with female users,
users completing graduate education, and with a worldwide user base
with a prominent number of visitors in the U.S., India, Spain and
Canada, among others.  In our own informal analysis of About.me users,
we find that they are often people who may benefit from having a
stable, open, public and visual presence on the Web for professional
reasons: creative, freelancing and marketing types are common job
profiles of About.me users.

While About.me users are clearly atypical Web users, they do
represent an important subset of OSN users that use multiple OSNs and
benefit from having a central point to aggregate their activities and
give an overarching biographical sketch of themselves.  We argue that
such users are important to study as they represent key aggregators
and disseminators of information on OSNs, by virtue of needing a
service like About.me to manage their distributed activities
and identity.  Aside from solving the user linkage problem manually
for us, About.me importantly 1) offers an application
programming interface (API) and 2) ensures that the data captured by
their site is publicly available, which are both considerations for
programmatic and reproducible analyses.

\begin{figure}[!tb]
\centering
\includegraphics[width=0.33\textwidth]{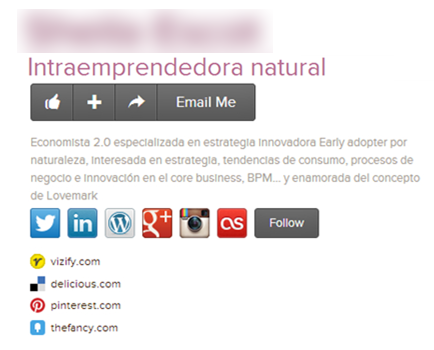}\vspace{-1mm}
\caption{A sample About.me profile linking to 10 of the user's websites.}\vspace{-1mm}
\label{fig_aboutmeprofile}
\end{figure}

\begin{figure}[!tb]
\centering
\subfigure[\# of users per OSN]{\includegraphics[width=0.44\textwidth]{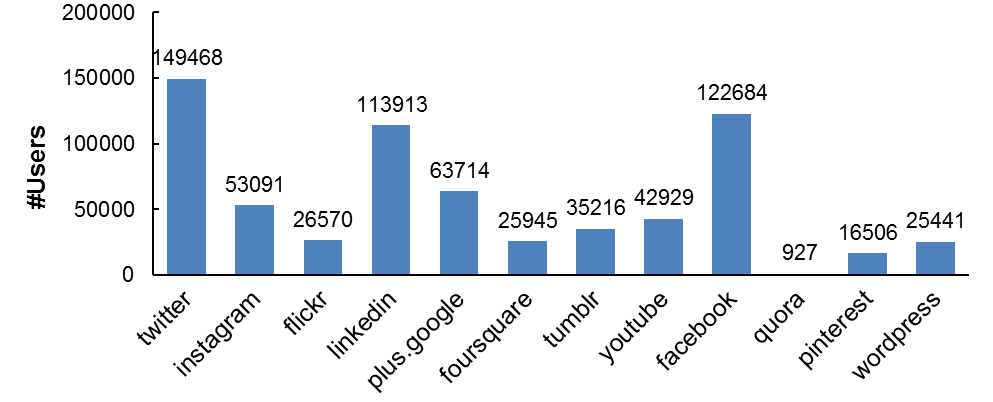}\label{fig_aboutmeCount_a}}
\subfigure[CDF of \# of accounts per user]{\includegraphics[width=0.44\textwidth]{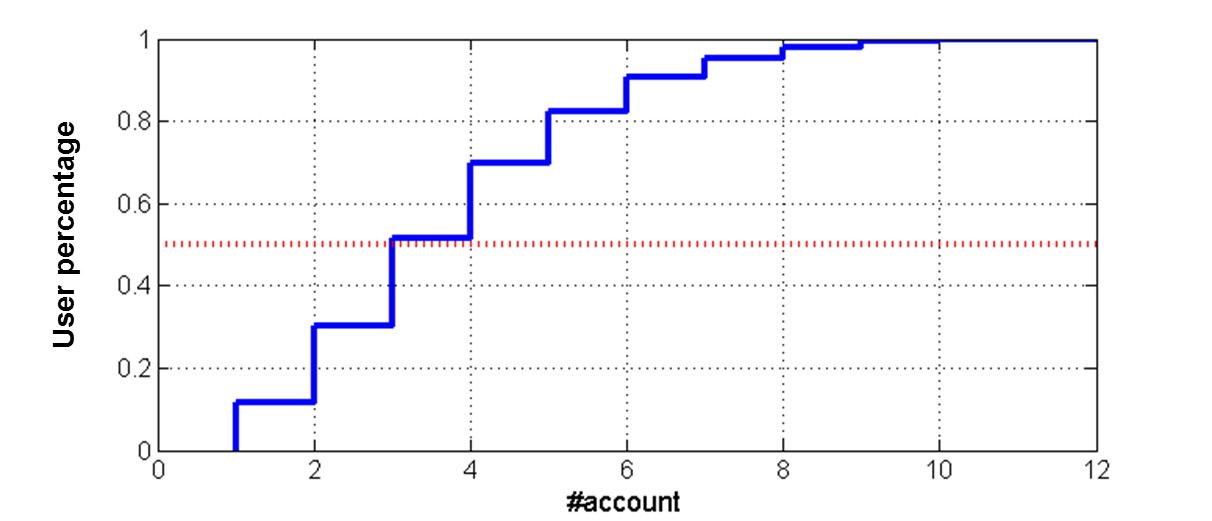}\label{fig_aboutmeCount_b}}
\vspace{-2mm}
\label{fig_aboutmeCount}
\caption{\# of About.me users linked to the 12 most frequently-linked OSN.}
\end{figure}

Using the About.me API, we collected a set of more than
$180,000$ registered user profiles.  Fig.~\ref{fig_aboutmeprofile} shows the profile
of Jane Doe (not her real name), which links to her OSN accounts and
websites (including Twitter, Linked.In and Wordpress).
From the cumulative distribution of the number of linked OSN accounts
per user, we see that the a slight majority of users list four or more
accounts (Fig.~\ref{fig_aboutmeCount_b}, red line).

%
We further limit the dataset used in this paper to users that link to
certain OSNs meeting three criteria for inclusion.  To ensure our
results are representative, we limit our study to the twelve most
frequently linked OSNs (Fig.~\ref{fig_aboutmeCount_a}).
We further require that the OSN expose most user information publicly
through an API, so that we could retrieve user activity; and that the
final OSNs chosen represent the breadth of media and functionality
common in today's Web~2.0 ecology.  With these selection criteria, we
selected the $15,595$ users that linked at least the average ($n=4$,
$\bar{n} = 3.7$) number of OSNs from the set of six OSNs: {\it
  Flickr}, {\it Google+}, {\it Instagram}, {\it Tumblr}, {\it Twitter}
and {\it YouTube}.  Here, we note that {\it Flickr}, {\it Instagram}
and {\it YouTube} focus on photo and video sharing, {\it Twitter} and
{\it Tumblr} are microblogging providers, and {\it Google+} is a
typical social networking site.
To compile the dataset used in this study, we crawl each user's
publicly accessible activities via the respective APIs on 15 August
2013.  Since all of the data we have obtained is public, and since we
believe that the compiled dataset is a valuable resource for studying
multiple OSN behavior, we have released our dataset for others to
conduct further
study\footnote{\url{http://to.be.made.public/if_accepted}.}.
Table~\ref{tab_aboutmebasic} gives statistics on the resultant
dataset.

\subsection{Statistics}

%
We first calculate the degree of overlap of users of one OSN with the others in our dataset using Jaccard similarity (Table~\ref{tab_newoverlap}).
Our figures are largely consistent with the previous Pew Internet
study that was performed over a global sample of online
adults~\cite{socialmedia2013}.  We see that Twitter has the largest
overlap, followed by Google+, revealing their popularities among
active social media users who interact with multiple OSNs.  Among
pairs, we see that YouTube shares $84.1\%$ users with Google+, likely
explained by their unique affiliation to Google and easy cross-sharing
mechanism; and that overlaps between Instagram and Tumblr are also
high ($78.8\%$ and $68.5\%$), which validates previous survey work
\cite{rainie2012photos} demonstrating that users prefer OSNs that
better support visual media.

\begin{table}
\centering
\caption{Demographics of our compiled About.me dataset.}
{
\begin{footnotesize}
\begin{tabular}{|l|r|l|r|}
\hline
{OSN}&{\# Users}&Activity Type&{\# Activities}\\
\hline\hline
{Twitter (Twi)}&{15,103}&microblog&{43,042,857}\\
\hline
{Google+ (G+)}&{12,445}&post&{2,522,873}\\
\hline
{Instagram (Ins)}&{11,922}&photo (upload)&{1,054,047}\\
\hline
{Tumblr (Tum)}&{10,259}&post&{8,171,592}\\
\hline
{Flickr (Fli)}&{10,139}&photo (upload)&{11,266,954}\\
\hline
{YouTube (YT)}&{8,883}&feed (upload)&{180,618}\\
\hline
\end{tabular}
\end{footnotesize}
}
\label{tab_aboutmebasic}
\end{table}\vspace{0mm}



\begin{table}
\centering
\caption{\% of users of one OSN who also participate in another OSN
  in our About.me dataset.}
{
\begin{footnotesize}
\begin{tabular}{|c|c|c|c|c|c|c|}
\hline
&\multicolumn{6}{c|}{also use}\\
\cline{2-7}
&{Twi}&{G+}&{Ins}&{Tum}&{Fli}&{YT}\\
\hline
\% of Twi&--&79.4&76.4&65.2&64.4&56.2\\
\hline
\% of G+&96.4 & -- & 73.5 & 61.7 & 61.0 & 65.0\\
\hline
\% of Ins& 96.7 & 76.8 & -- & 68.5 & 60.4 & 51.0\\
\hline
\% of Tum& 96.0 & 74.9 & 78.8 & -- & 59.4 & 49.2\\
\hline
\% of Fli& 96.0 & 74.8 & 71.0 & 60.1 & -- & 53.3\\
\hline
\% of YT& 95.5 & 84.1 & 68.4 & 56.6 & 60.9 & -- \\
\hline
\end{tabular}
\end{footnotesize}
}
\label{tab_newoverlap}
\end{table}

\begin{figure}[!tb]
\centering
\includegraphics[width=0.38\textwidth]{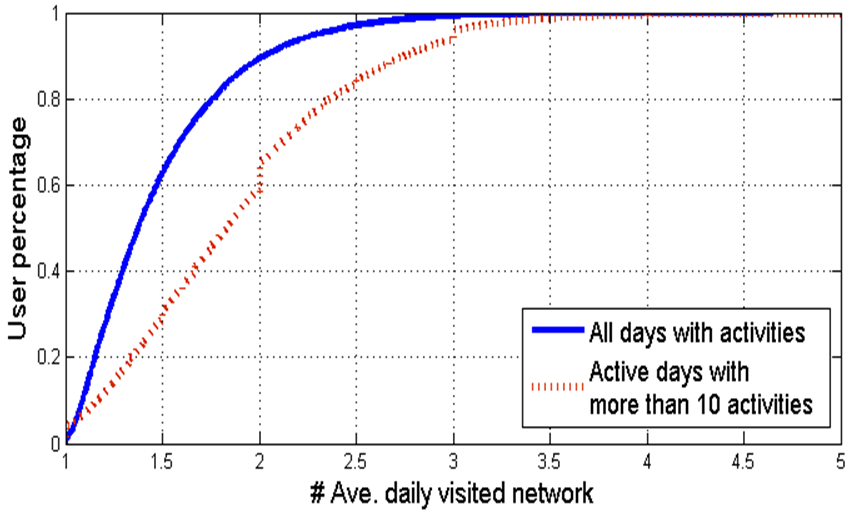}
\caption{Cumulative distribution of users' average daily OSN visits (excluding zeros).} 
\label{fig_cdfDaily}
\end{figure}

Merely linking an OSN account does not necessarily imply that the user
actively participates.  To gain deeper insight, we filter away dates
where users did not use one of the examined OSN.  We plot the average number of
networks visited daily as a cumulative distribution function 
(Fig.~\ref{fig_cdfDaily}), where a single user's different days of use
each contribute one data point.  We see that involvement on multiple
networks is quite common -- $\sim40$\% of all users in our dataset
are active on any given day, interacting with an average of $1.5$
networks.
We further examine that on highly active days -- {\it e.g.}, the days with over $10$ activities -- the average number of networks utilized is also correspondingly larger (dashed line in Fig.~\ref{fig_cdfDaily}).

%% file: 4_multi_network.tex
\section{Multi-network Participation Analysis}
\label{section:4}
Many online social network platforms have similar functions. Being able to follow individuals, post media and comment are pretty much ubiquitous activities that all OSN expose to users. Given this common functionality, why do users choose to use different OSNs? Is it due to their personal social networks (sharing things with different people who serve different social roles), or due to differences in functionality?
Our dataset also records users' self-reported profile description (on each network and About.me as well), in addition to historical activity logs of the users' interaction with each OSN. These two data sources allows the analyses to be done on each source, which can be used to triangulate support for conclusions and yield complementary perspectives. Our results from both source corroborate that different OSN platforms are utilised differently, and serve distinct purposes collectively.

\subsection{User Profiles}
Online social networks often encourage users to maintain and complete
their profiles, possibly to increase the user's vested interest in
using their network.  These optional profiles are typically short
sentences (e.g., a tagline) or paragraphs and are publicly accessible,
making them a good source of free text demographic information.  Here
we seek to answer the question: \textit{How do profile descriptions
  differ across multiple OSNs?}

\begin{table}
\caption{User profiles for two users who participate in all six
  OSNs. Nouns used in similarity calculation are bolded.}  {
\begin{small}
\begin{tabular}{|c|p{5.8cm}|}
\hline
\multicolumn{2}{|c|}{{\it User 1}} \\
\hline
Twitter&{\scriptsize{I'm a Digital \textbf{Media} \textbf{Specialist} passionate about self \textbf{education}, lifelong learning...}}\\
\hline
Google+&{\scriptsize{\textbf{Knowledge} is \textbf{freedom}. I run a \textbf{website} called DIY \textbf{Genius} that helps young \textbf{people} self \textbf{education}.}}\\
\hline
Instagram&{\scriptsize{Explore \textbf{Dream} Create.}}\\
\hline
Tumblr&{\scriptsize{I'm interested in digital \textbf{media}, \textbf{adventure} \textbf{sports}, and \textbf{mountains}.}}\\
\hline
Flickr&{\scriptsize{All my \textbf{photographs} are posted under the creative \textbf{commons} non commercial \textbf{attribution}...}}\\
\hline
YouTube&{\scriptsize{A \textbf{collection} of \textbf{videos} I've filmed on my \textbf{iPhone} while hiking \textbf{skiing} and \textbf{biking} in the \textbf{mountains}.}}\\
\hline
\hline
\multicolumn{2}{|c|}{{\it User 2}} \\
\hline
{\it All 6 networks} & {\scriptsize{\textbf{Web} \textbf{Geek}?, Senior Digital \textbf{Strategist} in \textbf{Melbourne}, social \textbf{media} \textbf{maven}, \textbf{Google}-aholic, \textbf{simplicity} and UX \textbf{advocate}...}} \\
\hline
\end{tabular}
\end{small}
}
\label{tab_self}
\end{table}

The user profile function is a common feature in OSNs, which is usually optional (as in the case of the studied six OSNs).
We expect that some users reuse the same textual description over all of their OSN accounts.  This is observed in our dataset (e.g., User~2) but actually make up a relatively small percentage of cases; of the users that filled all six OSNs profiles, 3\% used the same descriptions throughout.
More common were user profiles that hinted about the user's identity with respect to the common functionality of the OSN.  User~1 illustrates this case where each profile description is different, and customizes it towards the main functionality of the OSN; disclosing their job title in Google+ and Twitter, introducing personal interests in Tumblr, and summarizing their media contributions in Flickr and YouTube.

\begin{figure}[!tb]
\centering
\includegraphics[width=0.45\textwidth]{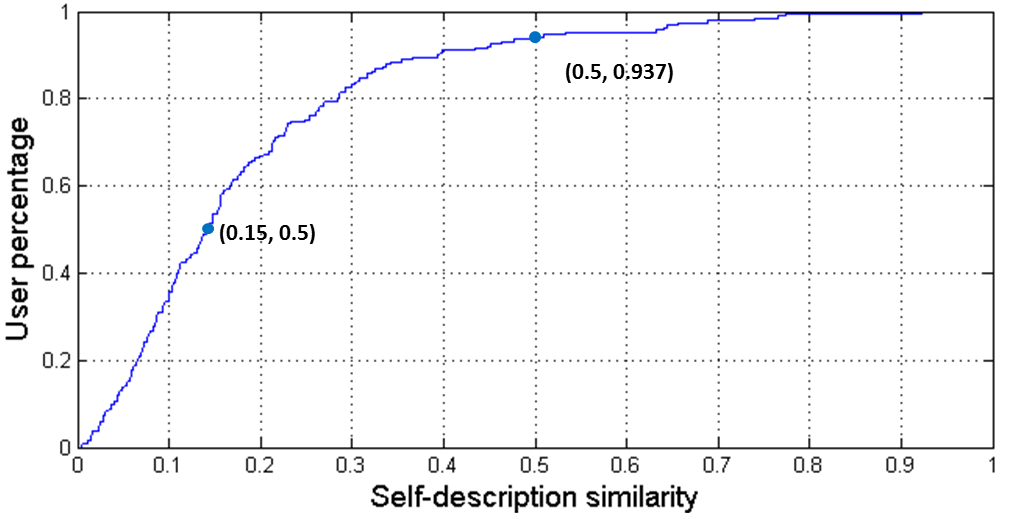}\vspace{-1mm}
\caption{Cumulative user distribution for self-description similarity.}\vspace{-1mm}
\label{fig_selfdescriptionsimilarity}
\end{figure}

To examine this issue further, we collected all users who had populated profiles in all six networks (inclusive of Users~1 and 2).  We further filtered out users whose descriptions were too short (less than 5 words), to lessen the impact of word sparsity.  On this final set of $190$ users, we first preprocess their profile descriptions in each network by retaining only frequent non-stopword nouns (as bolded in Table~\ref{tab_self}).  We employ the standard Jaccard coefficient over the two sets of remaining single-word nouns from the pairs of the user's profile descriptions, to calculate the average pairwise similarity of each user's profiles.

We plot the cumulative distribution of the average pairwise Jaccard profile description similarity in Fig.~\ref{fig_selfdescriptionsimilarity}.  We see most users have an average pairwise similarity significantly less than $0.5$ ($\sim95\%$, top labeled dot), which is considered a common threshold for sentential similarity in prior work~\cite{achananuparp2008evaluation}.  In fact, the majority users' profile similarity score is lower than $0.15$ (left labeled dot), indicating that most users describe themselves very differently across different OSNs.  Our follow-up manual sample analysis confirms this hypothesis.
We thus posit that social networks are distinct to users based on their functionality, and that the functional difference manifests itself in the users' self-description in the user profiles.

\subsection{Post Time}
Being habitual beings, timing describes the way we behave in our daily lives: activities such as eating, sleeping and working. Correspondingly, post timing gives us a way to ascertain how users behave on each OSN. In this subsection, we analyse user behaviour by looking at temporal metadata of all posts shared by the user. We study a larger pool of 27K About.me users that provide their locale information, from which we are able to find out the local time of each post. For this section, we ansswer the question: \textit{How does user sharing behaviour vary with time?}

\subsubsection{Time of Day Analysis}
\begin{figure}[h]
  \centering
  \subfigure[Weekend]{
  \includegraphics[scale=0.25]{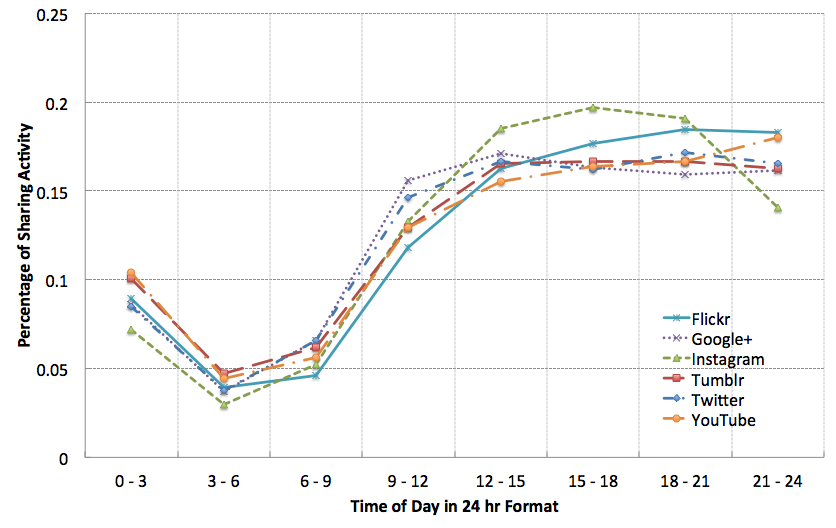}
  \label{fig:timeofdayweekend}
  }
  \subfigure[Weekday]{
  \includegraphics[scale=0.25]{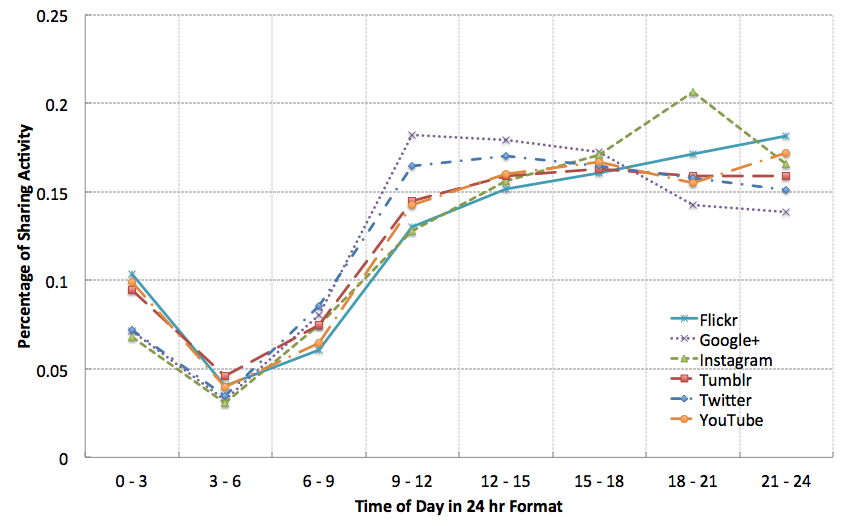}
  \label{fig:timeofdayweekday}
  }
  \caption{Distribution of user sharing activity over 24 hours for each social network}
  \label{fig:timeofday}
 \end{figure}

To begin, we analyse the time of day users are most active in sharing content on social networks.  We first divided the hours of a day into 8 intervals of 3 hours each and distinguishing between weekdays and weekends. For each user, we aggregate his/her sharing activity on each social network into a distribution over the 8 intervals. After summing and normalizing the distributions for all the users, we plot the resultant aggregated distributions as seen in Fig.~\ref{fig:timeofday}

Macroscopically, the results show that activity across all 6 social
networks follow a similar general trend. Low levels of activity are
observed in the wee hours of the night, and higher levels of activity
during the waking hours of the day and evening.

Taking a closer look, we notice that Google+ exhibits higher levels of activity during working hours (0900-1800) on weekdays, however, lower more evenly distributed levels of activity during and after working hours on weekends. Interestingly, Instagram shows an opposite trend, peaking in activity after working hours ($\textgreater$ 1800) on weekdays, and showing a decline after working hours on weekends. This hints, albeit subtly, at the contrasting nature of these two social networks -- Instagram is a platform more frequently used during  non working hours, while Google+ during working hours.

\subsubsection{Day of Week Analysis}
\begin{figure}[ht]
  \centering
  \includegraphics[scale=0.25]{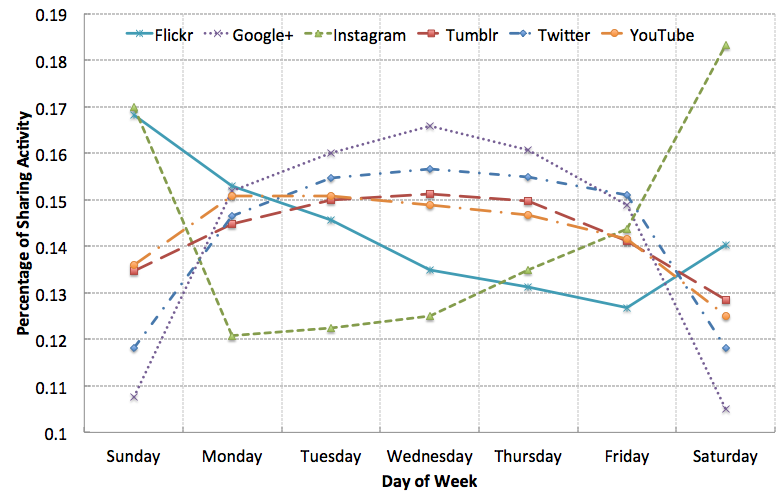}
  \caption{Distribution of user sharing activity over days of the week for each social network.}
  \label{fig:dayofweek}
\end{figure}

We see a clearer duality between OSN's usage during working times and non working times when we perform the same analysis over days of the week. Fig.~\ref{fig:dayofweek} shows the distribution of sharing activity over days of the week and provides us with several key insights. Image based social networks ({\it i.e.}, Flickr and Instagram) show a different trend throughout the week, peaking on the weekends, in comparison with the other video based, text and mixed media social networks that peak during the middle of the week - Wednesday.
One possible reason for this is that posts to these social networks are in the form of images, and users are less likely to be able to take photos because of work on weekdays.

To quantify the temporal similarity of the social networks in sharing activity,
 we employ Kullback-Leibler (KL) divergence to measure the difference between their posting day-of-week distribution. KL divergence is an asymmetric measure of difference between two probability distributions, 
 defined by the information loss when using one distribution to approximate another:
\begin{equation}
D_{kl} (P || Q) = \sum_{i}P(i) ln\frac{P(i)}{Q(i)}
\label{eq:kl}
\end{equation}

\begin{table}
\centering
\caption{KL Divergence scores between the distributions of OSN sharing activity over days of the week}
{
\begin{footnotesize}
\begin{tabular}{|c|c|c|c|c|c|c|}
\hline 
$Q$\textbackslash$P$ &\multicolumn{6}{c|}{KL Divergence Scores}\\
\cline{2-7}
&{Twi}&{G+}&{Ins}&{Tum}&{Fli}&{YT}\\
\hline
Twi & 0 & 0.0021 & 0.0365 & 0.0018 & 0.0172 & 0.0023\\                               
\hline
G+ & 0.0021 & 0 & 0.0226 & 0.0070 & 0.0075 & 0.0002\\
\hline
Ins & 0.0340 & 0.0217 & 0 & 0.0505 & 0.0131 & 0.0233\\                                    
\hline
Tum & 0.0019 & 0.0074 & 0.0553 & 0 & 0.0282 & 0.0072\\                                    
\hline
Fli & 0.0164 & 0.0073 & 0.0132 & 0.0263 & 0 & 0.0067\\                                    
\hline
YT & 0.0022 & 0.0002 & 0.0244 & 0.0069 & 0.0069 & 0\\          
\hline

\end{tabular}
\end{footnotesize}
}
\label{tab:daymatrix}
\end{table}

We chart a similarity matrix as seen in Fig. \ref{tab:daymatrix} using scores derived from Equation \ref{eq:kl} -- KL divergence for discrete distributions.
 We note how dissimilar Instagram and Google+ are. 
 Both of these networks are the extreme characterisations of the two general patterns of sharing behaviour OSNs observe over time: Most activity during the middle of the week, and contrastingly, most activity during weekends. 
 

\subsection{Profession}
As described in Section \ref{section:2}, About.me is a platform for users to advertise themselves professionally. Leveraging on this, we seek to answer the following question: \textit{Do users favour certain social networks for sharing personal and work related content?}

The separation of work related and personal posts are  user dependent, {\it e.g.}, photography is one's  hobby but another's profession.
Therefore, to answer the above question accurately, we need to  group users into professions.
We created sets of keywords that are related to 10 professions by identifying possible terms that adequately describe each profession. Then we search users' About.me profiles for these keywords and manually validate if each user to check if the keywords used are adequate in describing the profession. We prune and add more keywords and repeat this process until we are left with a set of keywords that we feel is suitable. These sets of keywords are used to select users for our profession group analysis.

In Table \ref{tab:professiondescription}, we provide example self-description for three 
  professions in About.me :
 \textit{Marketing Expert} (e.g., Musician, Podcaster, Filmmaker), {\it Producer}, and \textit{Developer}. We focus our analysis on these three types for two reasons: 1) 
 They are typical three professions in About.me, and 2) they are 
 easily distinguishable and thus help to simplify the identification and validation process.


\begin{table}[t]
	\caption{Profession description examples.}
	\label{tab:professiondescription}.
\begin{tabular}{p{40pt}p{180pt}}
\hline Profession & Self-description \\ 
\hline Marketing Expert &  {I am making new improvements new innovative solutions and new marketing tactics for e commerce affiliate marketing and vertical e commerce in Turkey. I want to make e commerce pie bigger. Also I am making consulting for e commerce and dotcom projects. If I believe in something, I have to do it.}\\ 
\hline Producer& I'm a Delhi based Electronic musician creative artist. I DJ produce minimal techno tracks as [...]. Also, I love robots. \\ 
\hline 
Developer & I'm a fifty something Scottish born software developer living in strand near cape town South Africa and working for [...] in Stellenbosch. I enjoy photography and am a committee member of [...] photographic society  photography. \\
\hline 
\end{tabular} 
\end{table}

Once knowing a user's profession, we then match his/her post's topic with the profession, and thus determine the post to be work related or personal.
To automate the process of inferencing topics from posts, we employ statistical topic modelling in the form of Latent Dirichlet Allocation (LDA) \cite{blei2003latent}. LDA models documents and their vocabulary in the same space, clustering similar documents and words together based on co-occurrence. Social Media posts being very short documents face the issue of insufficient verbosity for LDA to assign topics accurately. We tackle this issue by author-time pooling of posts as described in \cite{mehrotra2013improving}: Instead of a post being a single document, we treat a document as a collection of posts that have been shared within a predefined time interval by the user.
To construct our LDA model, we select a group of 2000 random users and train our model on a collection of 1,441,987 author-time pooled documents from their posts from all 6 social networks, with a predefined number of topics set to 50 by experimentation. We label each cluster that our model finds with a topic (if the top keywords that describe the cluster is coherent, otherwise we do not.) Next, we associate relevant topics to their respective profession groups.

\begin{figure}[t]
  \centering
  \includegraphics[scale=0.28]{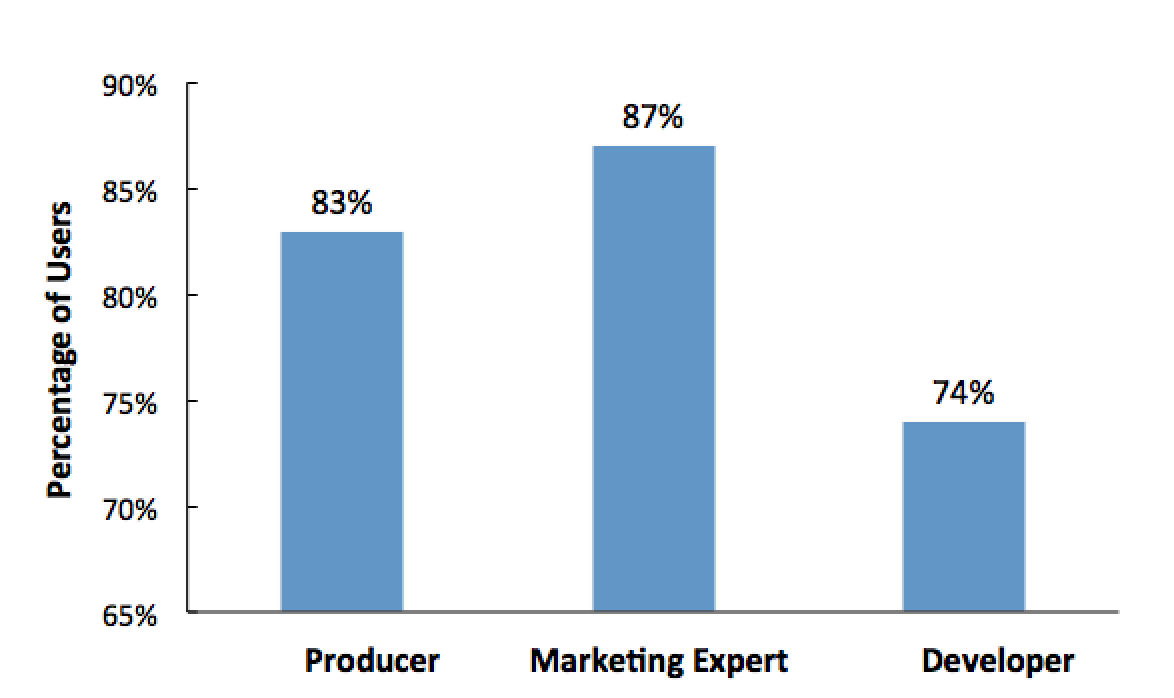}
  \caption{Percentage of users who frequently use social networks for professional use}
  \label{fig:professionuse}
\end{figure}

We start our analysis by asking: \textit{For each profession group, what percentage of users frequently share information about their profession (at all)?}. A user is considered someone who frequently shares information about their profession if at least one of the top 2 topics in any of the 6 social network matches that of the topics related to his or her job. Fig. \ref{fig:professionuse} shows a chart for the percentage of users in all 3 interests groups that do recently share information about their profession.

We start our analysis by asking: \textit{For each group, what percentage of users frequently share information about their profession (at all)?}. A user is considered someone who frequently shares information about their profession if at least one of the top 2 topics in any of the 6 social network matches that of the topics related to his or her job. Fig. \ref{fig:professionuse} shows a chart for the percentage of users in all 3 groups (of 100 randomly sampled users from each group) that frequently share content about their profession.

Developers do not frequently use social networks for professional use as compared to the other two professions. This phenomenon may be explained by the fact that the other two professions have more reasons to use OSNs as means of reaching out to their target audience: producers of content may use OSNs to publish their works and likewise, marketing experts use them as means of promoting their cause. However, there is no reason for developers to do the same, hence a less sharing of work-related content is observed for them.

\begin{figure}[t]
  \centering
  \includegraphics[scale=0.25]{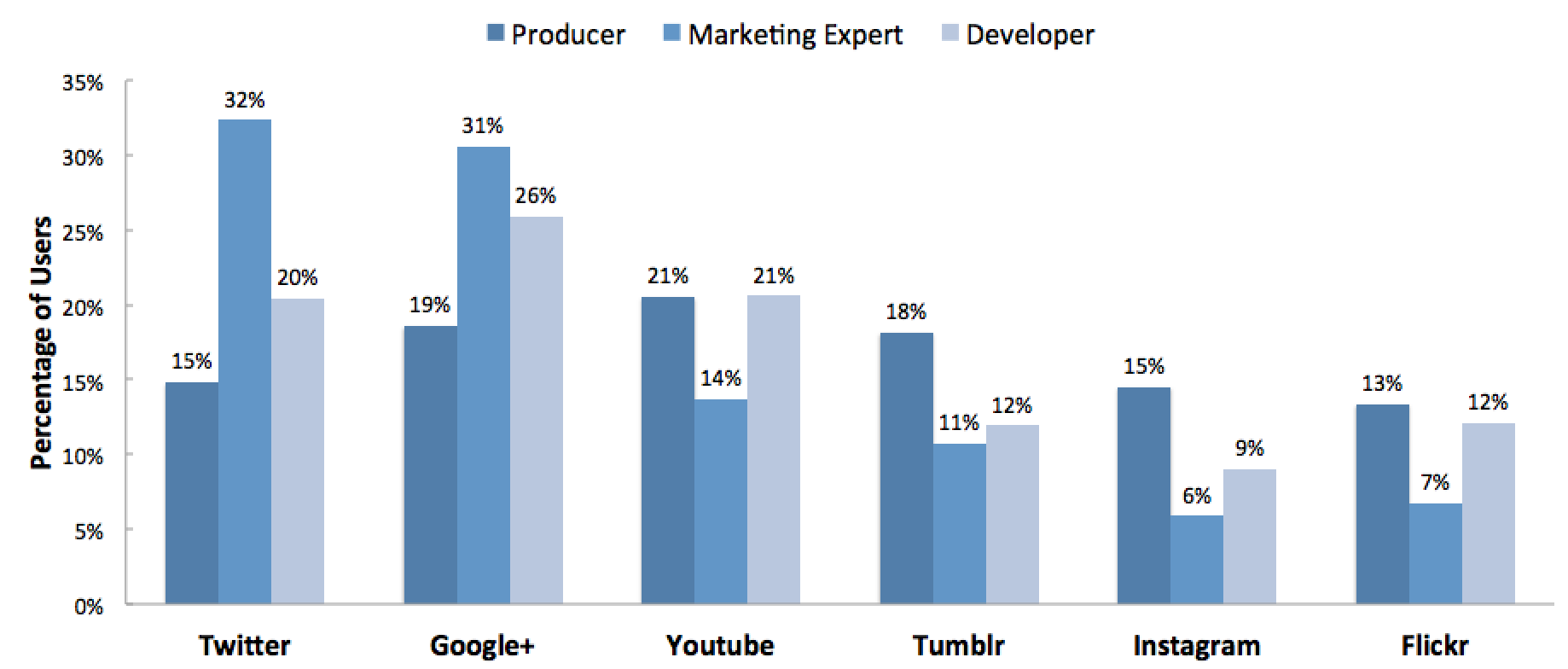}
  \caption{User's social network preference to share work related content}
  \label{fig:professionpreference}
\end{figure}
A natural follow up question is \textit{For those users that do share content related to their profession on social networks, which social networks do they prefer?}
Fig.~\ref{fig:professionpreference} shows that Google+ is consistently a popular choice amongst all three groups for sharing content related to work. YouTube is a video based OSN and 
is the primary choice for producer, 
as it is a suitable medium for them to distribute and publish their work.

For each profession, we investigate the most popular topics posted on each OSN for both hourly and daily intervals. We observe that the most popular topics that are posted to OSNs remain consistent over time (e.g. one instance of this observation is the most popular topic on Twitter for the Developer group is Technology for all days of the week) . This shows how each platform, to a user, has its own dedicated function and use i.e. a platform used mostly for work related purposes will unlikely be used for personal (non work related) use.

Our analysis in this subsection concurs with results obtained from our temporal analysis. Google+ sharing activity reaches its peak during working days and similarly, is the network of choice for professions that do not deal with media (music, video, images). Also, Instagram consistently ranks one of the lowest for all 3 professions supporting our earlier postulation of it being a network for personal rather than professional use. We can conclude from our analysis that some OSNs are used more frequently for work related purposes, whereas others are preferred for personal use.

%% file: 5_cross_network.tex
\section{Cross-network Interaction Analysis}
\label{section:5}
Informally, {\it cross OSN sharing} is when a user multicasts his
activity over multiple networks.  
For example, 
photos shared from Instagram to Flickr are automatically tagged as "uploaded:by=instagram", indicating their originality.
Like the Instagram/Flickr
case, cross sharing often has a source OSN platform (shared from) and
a sink (shared to).

To examine this user behavior, we build timelines for each user's
activities and identify cross-sharing activities.  An example timeline
fragment of one user is shown in Table~\ref{tab_usertimeline}, where we
observe cross-sharing.  Cross-sharing can be enabled by third party
software, which may broadcast the content simultaneously to target
OSNs.  We identify cross-shared (aggregate) activities as consecutive
activities on different OSNs, where common timing and content define
the bounds of the cross-shared activity (bolded instances in
Table~\ref{tab_usertimeline}).  In our dataset, we programmatically
identified $111,431$ cross-sharing activities.


Formally, we define cross-sharing as a $4$-tuple:
$$
CS:=<U,P,N^{source},N^{sink}>
$$
\noindent where $U$ is the user identification, $P$ is the post being cross-shared, $N^{source}$ indicates the network where the post is originally created (or {\it source}), and $N^{sink}$ indicates the destination network (or {\it sink}).
Note that we only assign source and sink roles only when evidence of the direction of sharing is present; e.g., Table~\ref{tab_usertimeline}'s the tweet's inlined shortened URL resolves to Instagram, and the photo on Flickr contains machine generated tag ``uploaded:by=instagram''.


\begin{table}[ht]
\centering
\caption{A fragment of a sample user's activity timeline, that illustrates a cross-shared activity (bolded).}
{
\begin{small}
\begin{tabular}{|p{1.3cm}|c|p{4.4cm}|}
\hline
{Time}&{Network}&{Content}\\
\hline\hline
{\footnotesize{26/05/2013 18:36:46}}&{Twitter}&{\scriptsize{my poolside jam... http://t.co/f2480xhw5u}}\\
\hline
{\footnotesize{27/05/2013 03:00:42}}&{\textbf{Instagram}}&{\scriptsize{I miss baby snuggles when the kids are away... :) I'm so blessed. \#momlife}}\\
\hline
{\footnotesize{27/05/2013 03:00:44}}&{\textbf{Flickr}}&{\scriptsize{I miss baby snuggles when the kids are away... :) I'm so blessed. \#momlife}}\\
\hline
{\footnotesize{27/05/2013 03:00:44}}&{\textbf{Twitter}}&{\scriptsize{I miss baby snuggles when the kids are away... :) I'm so blessed. \#momlife @ home tweet home http://t.co/WioRNR6BjA}}\\
\hline
{\footnotesize{27/05/2013 21:02:02}}&{Twitter}&{\scriptsize{you just never know who or what's going to show up at a wedding ... ;)}}\\
\hline
\end{tabular}
\end{small}
}
\label{tab_usertimeline}
\end{table}

Given the identification of these cross-sharing activities, we
characterize cross-sharing behavior in terms of platform
considerations in the following.

\begin{figure}[!tb]
\centering
\includegraphics[width=0.43\textwidth]{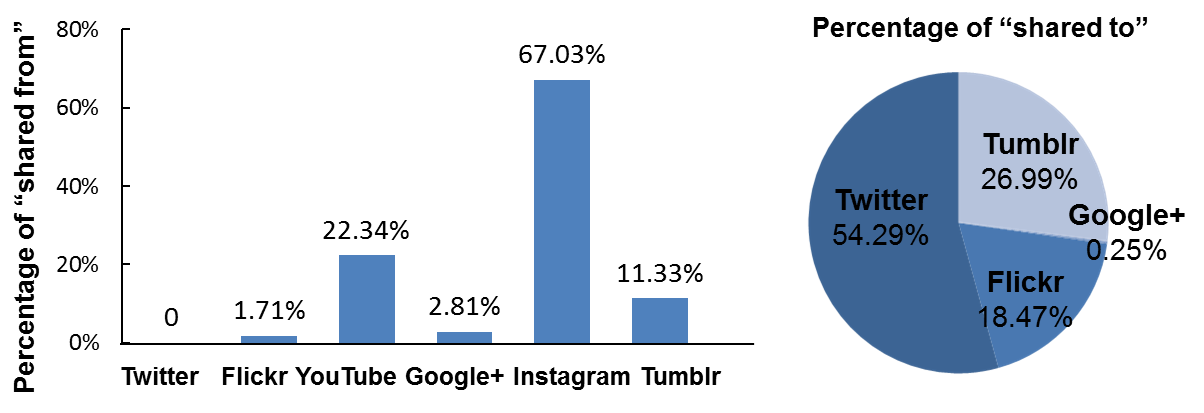}\vspace{-1mm}
\caption{Cross-sharing per OSN. (l) percentage of posts {\it shared
    from} an OSN; (r) percentage of all identified cross-shared
  activities in our dataset {\it shared to} a particular
  OSN.}
\label{fig_bereshare}
\end{figure}
\begin{figure}
\centering
\includegraphics[width=0.42\textwidth]{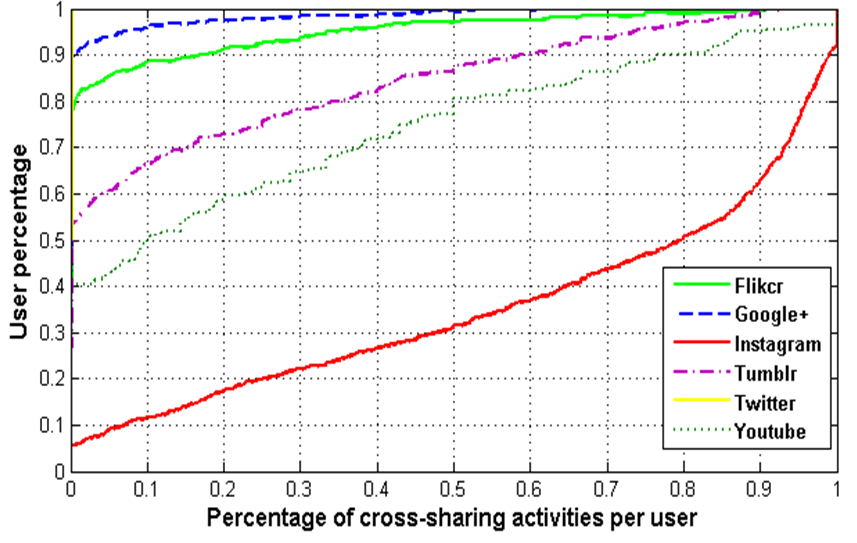}\vspace{-1mm}
\caption{Cumulative distribution of users with respect to their percentage of cross-sharing activities.}
\label{fig_crosssharePerSite}
\end{figure}

We map the dissemination flow of user content in the six OSNs by
aggregating the respective $N^{source}$ and $N^{sink}$ identities over
all detected cross-sharing activities.
Fig.~\ref{fig_bereshare} (left) shows the percentage of all posts in each OSN that is cross-shared to another OSN.
Instagram and YouTube hold the highest {\it shared from} percentage and serve as the most significant source OSNs.
We also plot (in Fig.~\ref{fig_crosssharePerSite}) the CDF of users with respect to
their cross-sharing ratio; i.e., the percentage of the user's own
content that are also disseminated to another OSN.
Cross-sharing turns out to be platform-sensitive.
Instagram serves as a popular source; over $90\%$ of Instagram users have shared their Instagram posts to another network.
For Flickr and Google+, fewer than $20\%$ of users took their original platform content and shared it to another OSN.

We graph each platform's share of all 111K+ identified cross-shared activities as a destination (sink) in Fig.~\ref{fig_bereshare}(r).
Twitter dominates, being the dominant destination for $54\%$ of cross-sharing activities.
We believe that OSNs examined play largely different roles in source/sink discrimination.
Source and sink networks for cross-sharing activity are markedly different.



\begin{figure}
\centering
\includegraphics[width=0.23\textwidth]{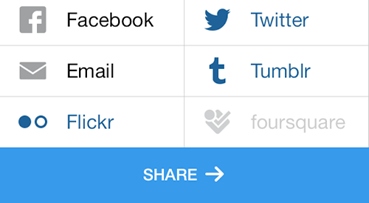}\vspace{-1mm}
\caption{Sharing functionality in the Instagram mobile application.}\vspace{-1mm}
\label{fig_instagramshare}
\end{figure}

\begin{figure}[!tb]
\centering
\includegraphics[width=0.45\textwidth]{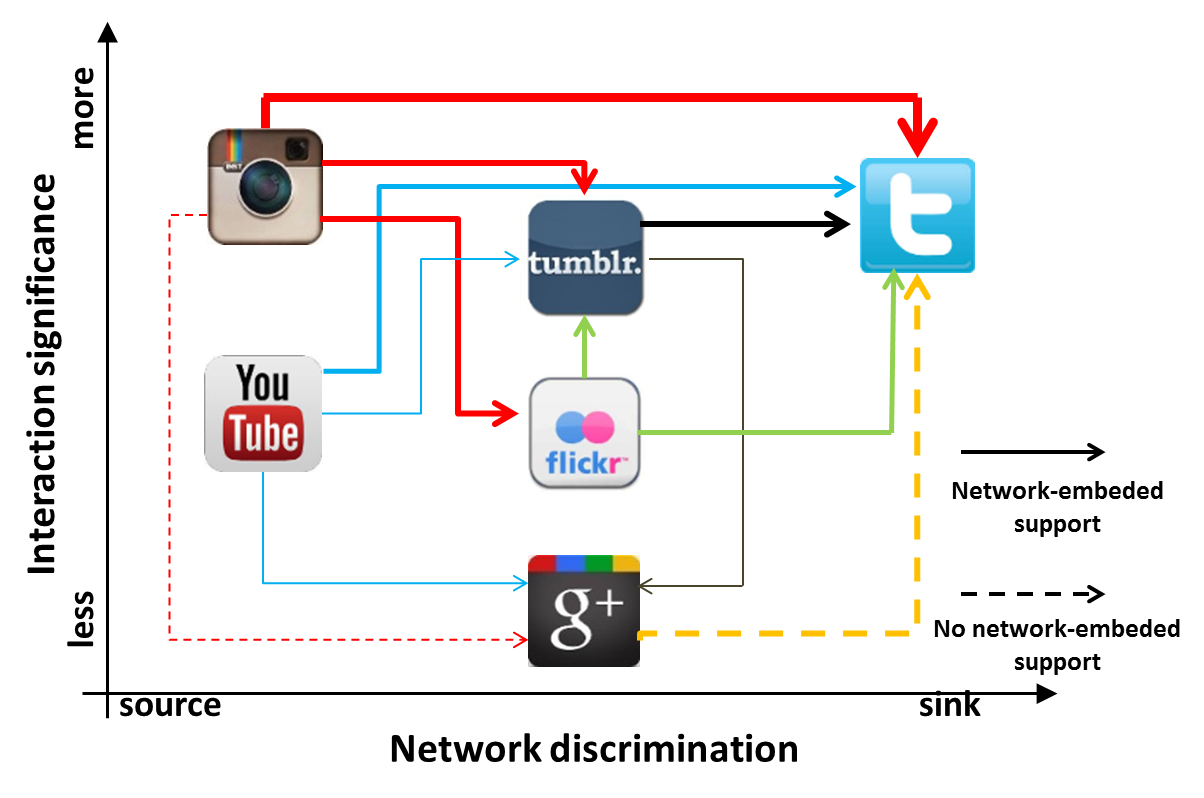}
\caption{User generated content flow among OSN platforms. Content flows in the directions indicated by arrows and line thickness represents flow volume.  Arrow color correspond to a specific OSN.}
\label{fig_flowmap}
\end{figure}

Instagram user content originates from its mobile application that places sharing as a central theme, manifesting in the embedded ``Share'' button to route the post to various social media networks.
As shown in Fig.~\ref{fig_instagramshare}, the central placement of the sharing functionality and its usability (dedicated buttons for different OSNs) drives the cross-sharing behavior we observe.
This suggests that platform functionality strongly drives cross-sharing.
We examined the current ({\it ca.} April 2015) state of embedded cross-sharing options for each of the six networks, both in their website versions as well as mobile apps for the iOS and Android mobile platforms, presented in Table~\ref{tab_crossService}.
We find that Instagram and YouTube support the most sharing destinations; Tumblr and Flickr support sharing to certain sinks, while Twitter and Google+ do not support sharing.  Aside from in-platform support, cross-sharing can be done through third-party plug-ins, (e.g., friendplus.me for Google+) or done by the user manually.
%
Fig.~\ref{fig_flowmap} illustrates our understanding of the information flow among the networks examined, where the horizontal axis represents the tendency of a platform to serve as source or sink, and the vertical axis represents the
level of explicit cross-sharing interaction.
With some caveats, we see that OSNs that focus more on visual media tend to be sources, whereas platforms focusing textual media serve more as sinks.  Tumblr, Flickr and Google+, positioned in the middle, play both roles.

We make several additional observations surrounding cross-sharing source (source support). Although Instagram cut off its sharing to Google+ after being acquired by Facebook (April 2012), some of the About.me users in our dataset manually cross-share their Instagram photos to Google+.  Also, while Google+ positions itself as a sink network, and does not support any cross-sharing, users have found means to do so through other workflows.
We believe both source and sink characteristics have their roles within the OSN ecology: ``Sourceness'' acknowledges original content and possibly the value add of using a particular platform (artistic filters in Instagram may be an example), while ``sinkness'' promotes an aggregator effect that enables downstream analytics to gain more complete pictures.

\begin{table}[!tb]
\centering
\caption{Cross-sharing support by platform.}
{
\begin{small}
\begin{tabular}{|c|p{5.4cm}|}
\hline
{}&{Supported destination network}\\
\hline\hline
{Twitter}&{--}\\
\hline
{Google+}&{--}\\
\hline
{Instagram}&{Twitter, Flickr, Tumblr}\\
\hline
{Tumblr}&{{\it depends on the user's personalized setting}}\\
\hline
{Flickr}&{Twitter, Tumblr}\\
\hline
{YouTube}&{Twitter, Tumblr, Google+}\\
\hline
\end{tabular}
\end{small}
}
\label{tab_crossService}
\end{table}

%% file: 6_conclusion.tex
\section{Conclusion}
\label{section:6}

We studied multiple online social networks (OSNs) from a user-centric
perspective, with the aim of discovering behavioral patterns in their
multiple network use.  We analyzed how users participate and interact
among six popular OSNs in our crawled dataset which we have made
publicly available.  Our study validates the hypothesis that users
exhibit varied behavior on different OSNs, accounting in part for the
OSN's primary media type.  In our multi-network analysis of single
OSNs, we initially showed how the majority of users portray themselves
differently across OSNs, suggesting differences in use. By examining
users' sharing activities, we uncovered a dichotomy between usage for
professional and personal reasons.  In our cross-network sharing
analyses, we mapped how users post from one source network to a sink
network.  By plotting the source--sink directionality of
cross-sharing, we labeled the media-centric OSNs of YouTube and
Instagram to be sources, and the lowest common denominator Twitter OSN
to be the common sink.

Our study has examined the public face of OSNs, uncovering just the
surface of the vibrant and varied ecology that is today's social
network.  While our study covers a much larger scale than previous
works that have largely confined themselves to the analyses of one or
two individual network, a key limitation of our work is that we have
only studied largely networks, and their user's public sharing
activities.  In the future, we plan to harvest data from private
social networks where possible, to further gain insight on the
differentiation between public and private OSN use.  We believe such
work may benefit improve social media applications.